# Inelastic Scatterings of Entangled Mössbauer Gammas

Yao Cheng, Zhongming Wang
Department of Engineering Physics, Tsinghua University, Beijing 100084, China
E-mail: yao@tsinghua.edu.cn

We report the observation of the temperature-dependent inelastic scattering of three entangled Mössbauer gammas in the time-resolved Mössbauer spectroscopy. Recently, the long-lived E3 Mössbauer transition of rhodium generated by bremsstrahlung irradiation has been reported. Two kinds of X-rays with the fast decay are attributed to the tri-photon effect. They are tri-photon pile-up of rhodium K X-rays and the high-Z impurity K X-rays. Energy of the particular K emission is higher than the sum energy of two Mössbauer gammas. This letter reports new discoveries by cooling down the sample using liquid nitrogen, namely the collective anomalous emission of entangled Mössbauer gammas. The enhancement of inelastic scatterings at low temperature such as rhodium K satellites is attributed to this entanglement.

## 1. Previous Observations

In the previous papers we reported the observation of long-lived rhodium Mössbauer effect generated from bremsstrahlung irradiation [1,2]. In this letter, the Mössbauer effect with extremely narrow linewidth of $10^{-19}$ eV is further demonstrated by cooling down the sample temperature. Abnormal counts near 65 keV were previously attributed to the tri-photon K X-rays characterized by the fast decay [1]. We have realized further that the reported tri-photon effect contains not only the tri-photon pile-up but also single-photon counts. The single-photon counts are characteristic K of the high-Z impurities, whereas the rhodium tri-K counts are widely spread due to the tail pile-up by photon bunching. Impurities such as Bismuth excited by this particular tri-photon effect have been identified by applying extra magnetic field. The tri-photon K X-rays at 65.1 keV and 66.8 keV are elevated by Kα internal conversions of the 4.02-day impurity $^{195m}$Pt transition, which was excited by the bremsstrahlung irradiation and led to the wrong conclusion of energy shift to the low-energy side [1].

## 2. Tri-γ Interpretation

We suggest in this letter that the tri-photon effect is attributed to three Mössbauer gammas. Each gamma propagates like the multi-beam nuclear Borrmann mode [3]. Borrmann mode is the known X-ray eigenmode in crystal from field cancellation at the lattice sites, and thus the photo-electric effect is suppressed [4]. In the mean time, nuclear coupling is possible and even enhanced or suppressed for the multi-polar nuclear transition [3]. Matching condition in the crystal lattice provides the long-range Borrmann channel for the coherent superposition of three gammas with highly correlated directionality, polarization, and phase. This is similar to the entangled bi-photon state generated by spontaneous parametric down conversion [5]. The previously reported tri-photon effect can be described as the particular phenomenon of three entangled gammas (hereinafter tri-γ). In this letter, we shall provide evidences of this tri-γ interpretation. Since the Borrmann effect is temperature dependent, it becomes more significant at low temperature due to the reduction of atomic vibration. When atoms deviate from their equilibrium position, they become a scattering center of the tri-γ. Consequently, the impurities are also the scattering centers of the tri-γ. The inelastic scattering of impurities with binding energies (Bi) greater than 80 keV reveals that the reported nuclear Borrmann mode is not a single-photon effect but an entangled tri-photon effect [1].

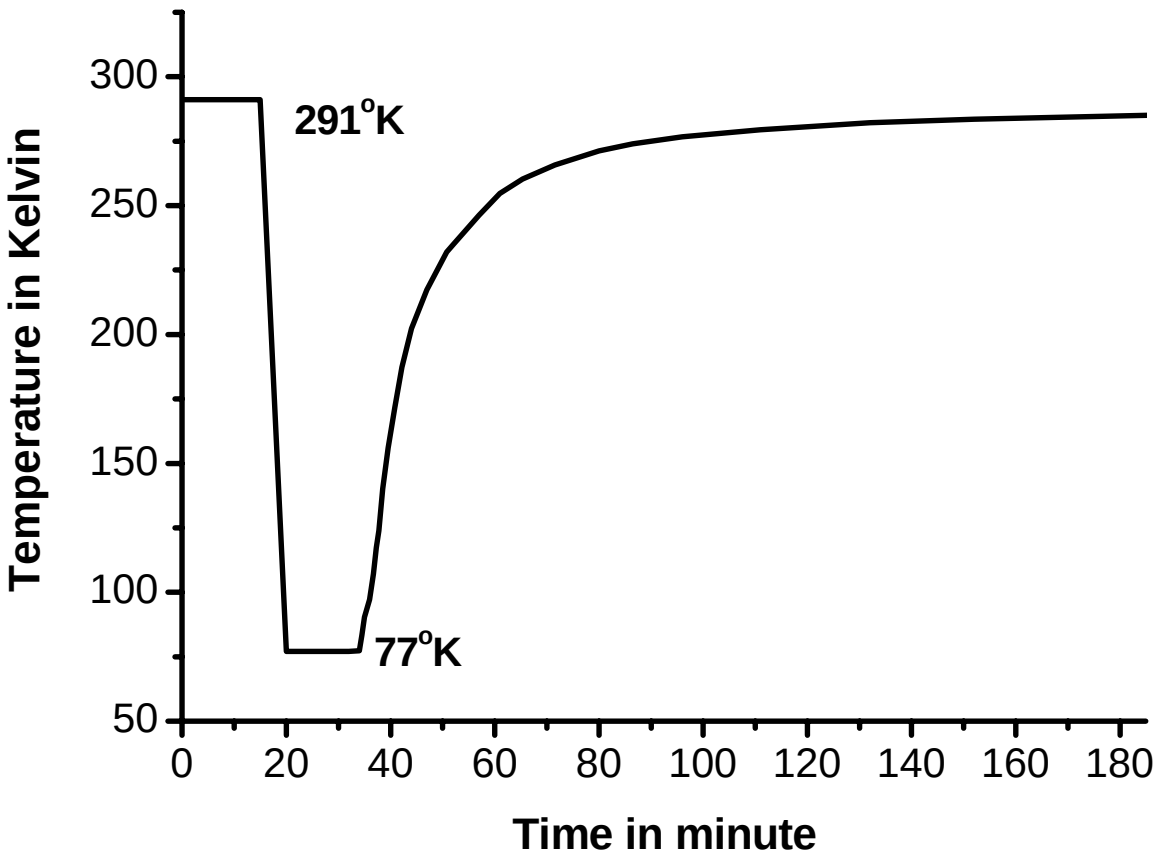

FIG. 1: Typical time evolution of the sample temperature measured with Pt 1000. The sample was kept at room temperature of 18°C for twenty minutes and then immersed in liquid nitrogen for twelve minutes.

## 3. Experimental Conditions

We have applied 120-minute irradiation with the same excitation procedure reported previously [1,2]. Transportation delay of several minutes between irradiation positions and detector is negligible to the long-lived rhodium isomer. Measurements were carried out with the leveled HPGe detector as illustrated in Figure 2b of reference 2, instead the N-S detecting direction is applied. Now, we used a smaller HPGe detector of CANBERRA GL0210P with an active area of ϕ-1.6 cm. Distance between the detector and the sample was 1 cm. The tri-photon counts [1] become less significant due to the small collimation angle of this experimental arrangement.

The square sample had a dimension of 2.5 cm × 2.5 cm × 1 mm with 99.9% purity of rhodium (Rh00300, Goodfellow) [2]. The rhodium sample is a poly crystalline with the fcc lattice structure. The irradiated sample was kept in room temperature at 18°C for twenty minutes and then liquid nitrogen was poured in to immerse the sample from one side for twelve minutes. Liquid nitrogen was able to overflow the sample to the other side from time to time, and indeed repetitive double-side quenching was created. The sample temperature recovered back within one hour. The typical temperature behavior is measured later as shown in Fig. 1.

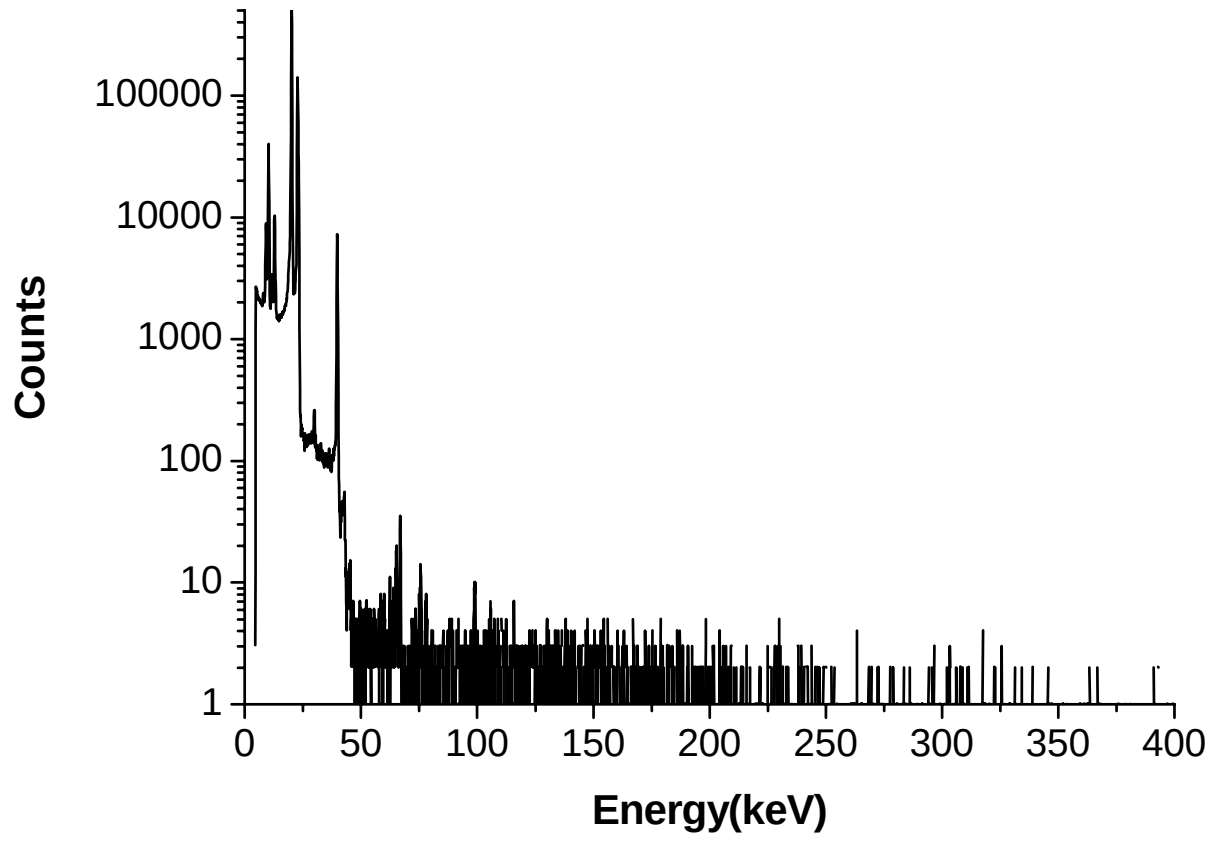

FIG. 2: Energy-resolved spectrum of rhodium isomeric emissions collected in 4096 channels within three hours. Around 20 keV are the K internal conversions. Their escape peaks locate around 10 keV. Gamma is at 39.8 keV. Three peak pile-ups of K lines are next to the gamma at the right-hand side.

## 4. Observations

Figure 2 illustrates the spectrum accumulated in three hours. The Kα lines at 65.1 keV and 66.8 keV of the $^{195m}$Pt transition and its gamma at 98.8 keV are identified. The total initial count rate was about 700 cps mainly contributed by the rhodium K lines. The estimated initial pile-up rate of gamma and K is 0.03 cps between 46 keV and 63 keV due to the 3-μs shaping time of ORTEC 572 amplifier, assuming that the pile-up of three K X-rays by accident is insignificant in this region. We observed ~0.1 cps at the beginning of Fig. 3. This large count rate is suppressed by turning on the pile-up rejection. We have inserted copper absorber sheets between sample and detector one by one to check the peak pile-up of K lines. The pile-up rates are proportional $p^2(1-p)$ instead of $p^2$, where p is the probability to detect a K photon. The anomalous tail pile-up is thus attributed to tri-K photon bunching. While the initial pile-up rate by accident is $10^{-4}$ cps in the region between 63 keV and 90 keV, we shall not observe any decay in figure 4, except the constant term contributed from the $^{195m}$Pt transition and lead shielding. Particular observations of interest are the count enhancement after the quenching in figures 3 and 4. We conjecture that they are due to the inelastic scattering of tri-γ.

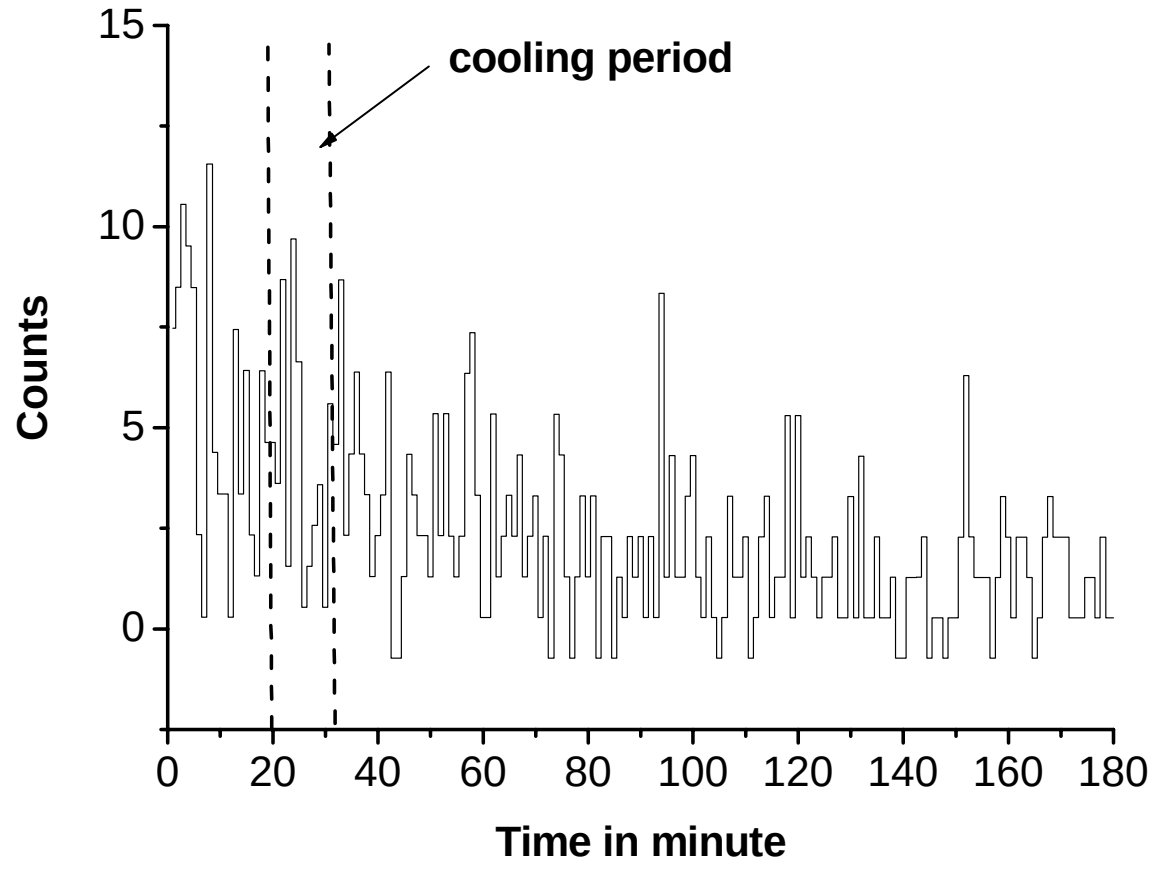

FIG. 3: Time evolution of the pile-up counts between 46 keV and 63 keV. Each data collection period is one minute. The average background has been removed. We started the quenching preparation one minute before the illustrated cooling period.

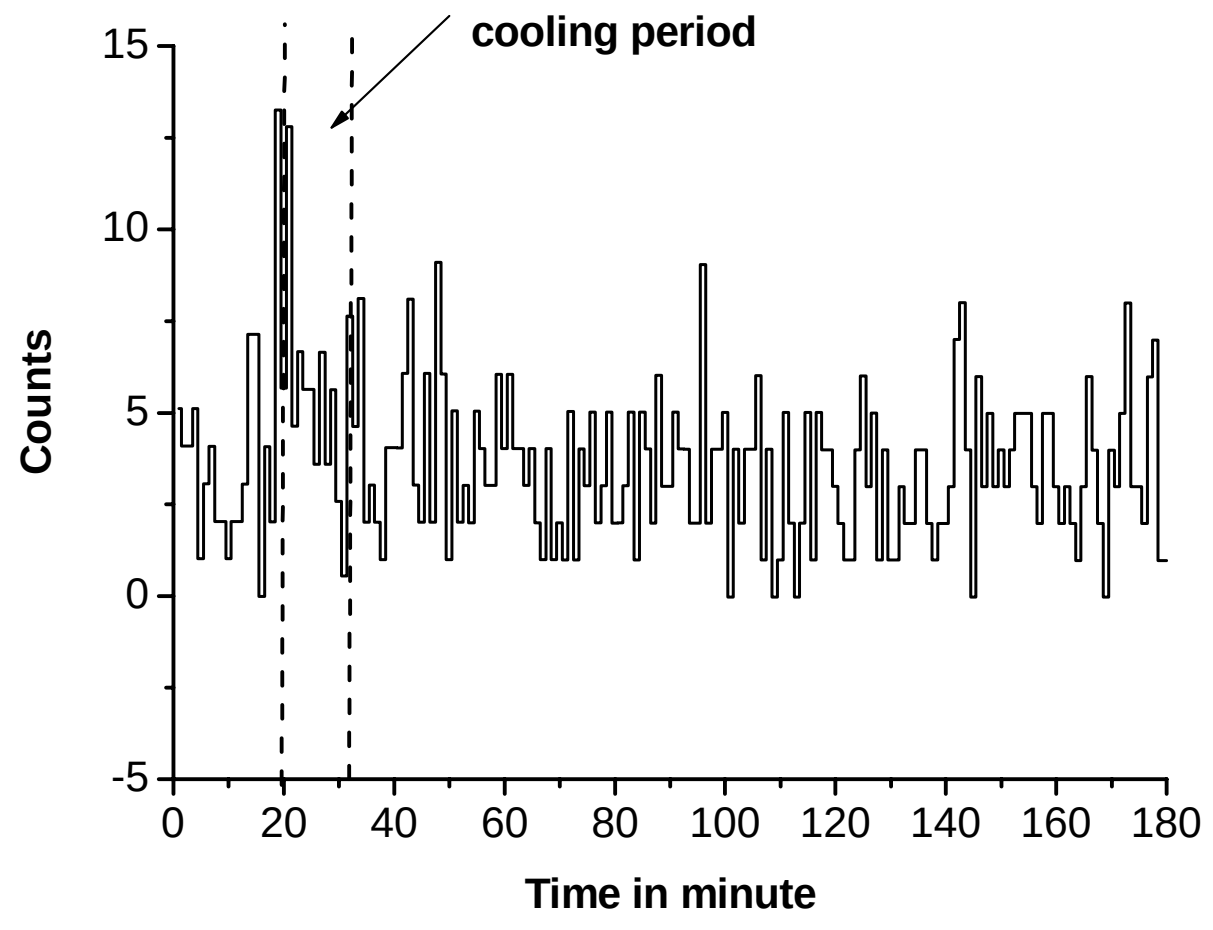

FIG. 4: Time evolution of the tri-γ scatterings between 63 keV and 90 keV. Each data collection period is one minute. The average background was subtracted. We started the cooling preparation at 19th minute that might lead to the count increase.

Figure 5 illustrates the time evolution of K lines and gamma. The speed-up decay during the cooling period was obviously observed. Figure 6 explores the Kα lines at 20.2 keV from the peaking channel 182 to the upper channel 184. Each channel has a 0.1-keV bandwidth. Three channels have different behaviors. Their ratios in Fig. 7 demonstrate the details of K-satellite excitations [6]. The K-satellite counts at upper Kα channels increase slightly during the cooling period. We have checked the double-Gaussian K peaks using the energy resolved spectrum. Both Kα and Kβ lines shift 7 eV in average to the right hand side (higher energy) by cooling. The ratio between $K\alpha_2$ and $K\alpha_1$ deviates from the tabled value of 53 to a lower value of 30, whereas the ratio between $K\beta_2$ and $K\beta_{1,3}$ has no significant change referred to the tabled value. This ratio $K\alpha_2/K\alpha_1$ remains almost constant during the three-hour measurement but nearly 10% increased during the cooling period. This ratio change reveals the enhanced multi-polar scattering due to coincidence creation of $L_3$ hole and K hole. Figures 8 and 9 illustrate the ratios of Kα/Kβ and K/γ, which demonstrate the nitrogen absorption and suppression of K emission. In the first six minutes of cooling, most of the liquid nitrogen between sample and detector is evaporated. No significant absorption is observed in this time period, and thus the reduction of the internal conversion K/γ can not be attributed to the nitrogen absorption but the suppression of K emission. The slow increment of K/γ reveals the speed-up of gamma decay [2].

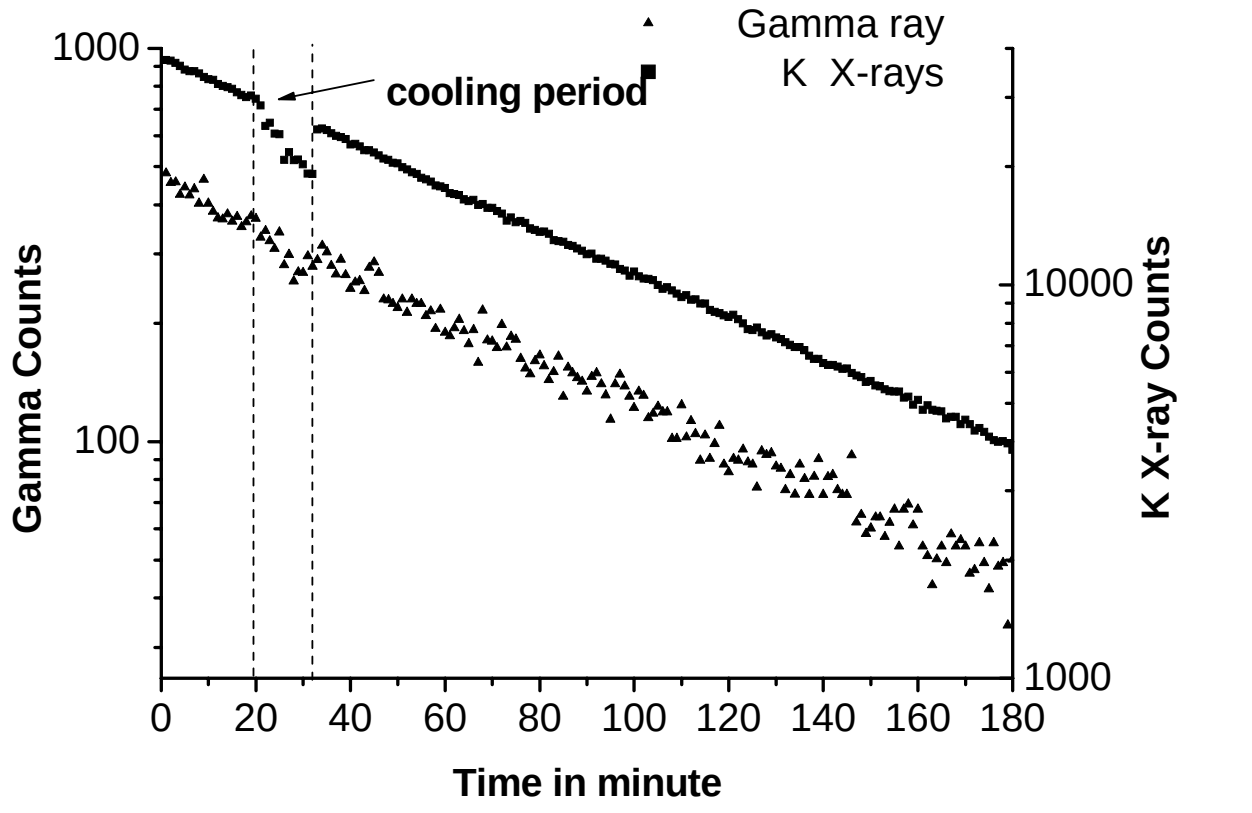

FIG. 5: Time evolution of the K lines and gamma emitted from the long-lived isomeric state. The K lines are collected from 35 channels between 20 keV and 23.5 keV. Gamma is collected from 16 channels centered at 39.8 keV. The right ordinate is the K counts and the left ordinate is the gamma counts.

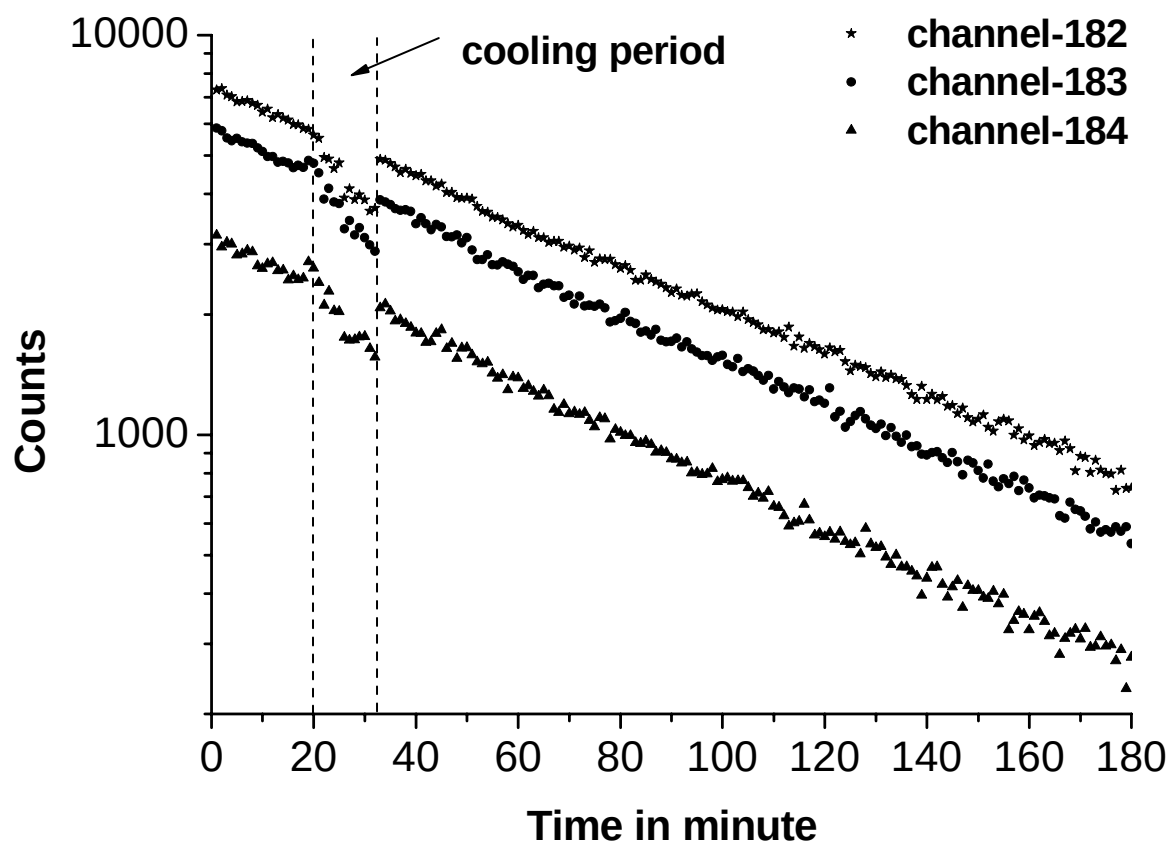

FIG. 6: Time evolution of the Kα lines in details. Channel 182 is the 0.1-keV peaking channel at 20.2 keV. Channel 183 and 184 are the upper channels.

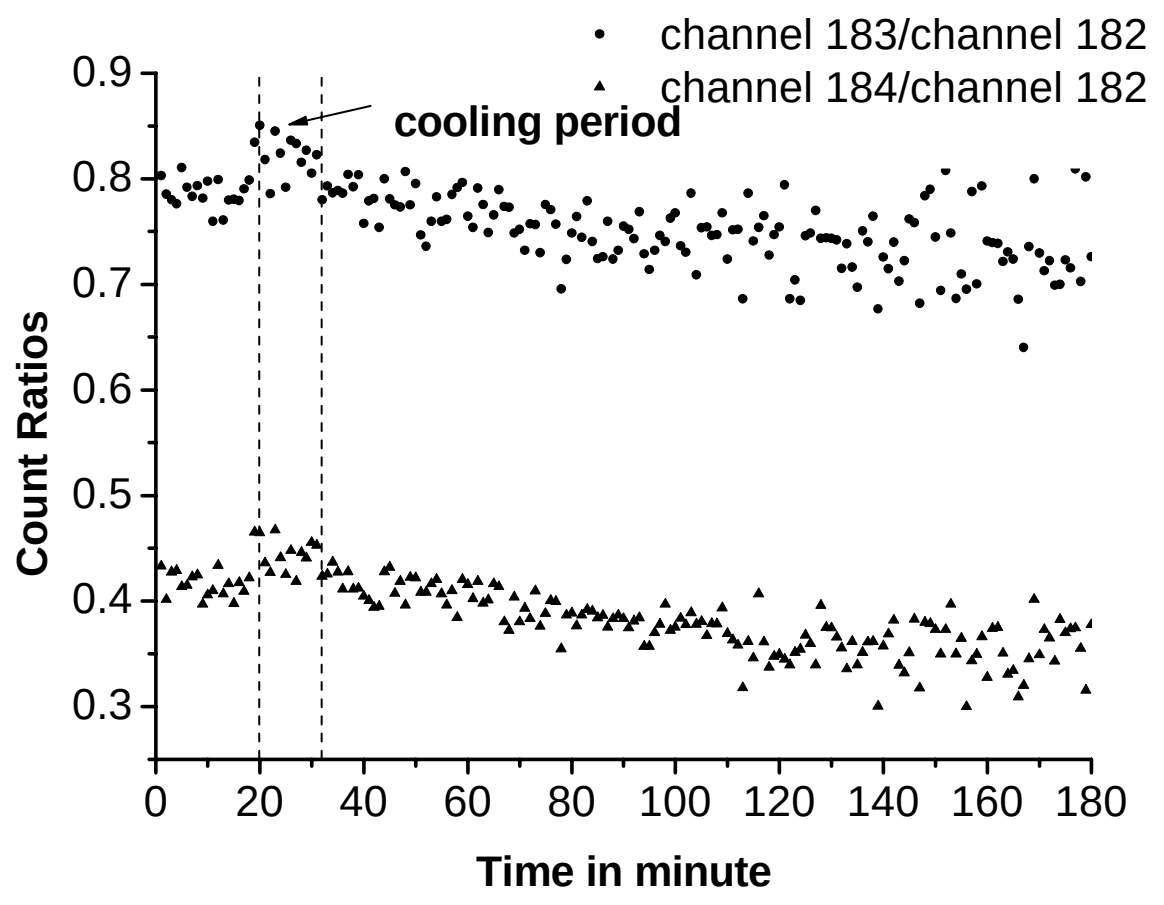

FIG. 7: Time evolution of the Kα satellites, which are presented in channels 183 and 184 and are normalized by the count of the 182 peaking channel at 20.2 keV. Slow drift downwards is due to baseline shift.

## 5. Conclusions

The speed-up rhodium decay at low temperature is a further evidence for the long-lived Mössbauer effect. Averaged K reduction is about 15% during the cooling period. The contribution from absorption of the overflowed liquid nitrogen is less than 7% obtained from the Kα/Kβ ratio. This K suppression reveals the coherent enhancement [7] of tri-γ or single gamma. Figure 7 of the satellite shift strongly supports the validity of the tri-γ enhancement at low temperature. The tri-γ interpretation reported here is attributed to the collective anomalous emission from the active internal Mössbauer source of E3 multipolarity [7].

This Hannon-Trammell anomalous-emission effect will be enhanced by lowering the sample temperature. Higher entanglement of 3n gammas is thus possible, while m≠3n entanglement leads to symmetry breaking in the (1,1,1) Borrmann channel. Although we did not observe the tri-γ directly, these are convincing evidences point to this tri-γ mode.

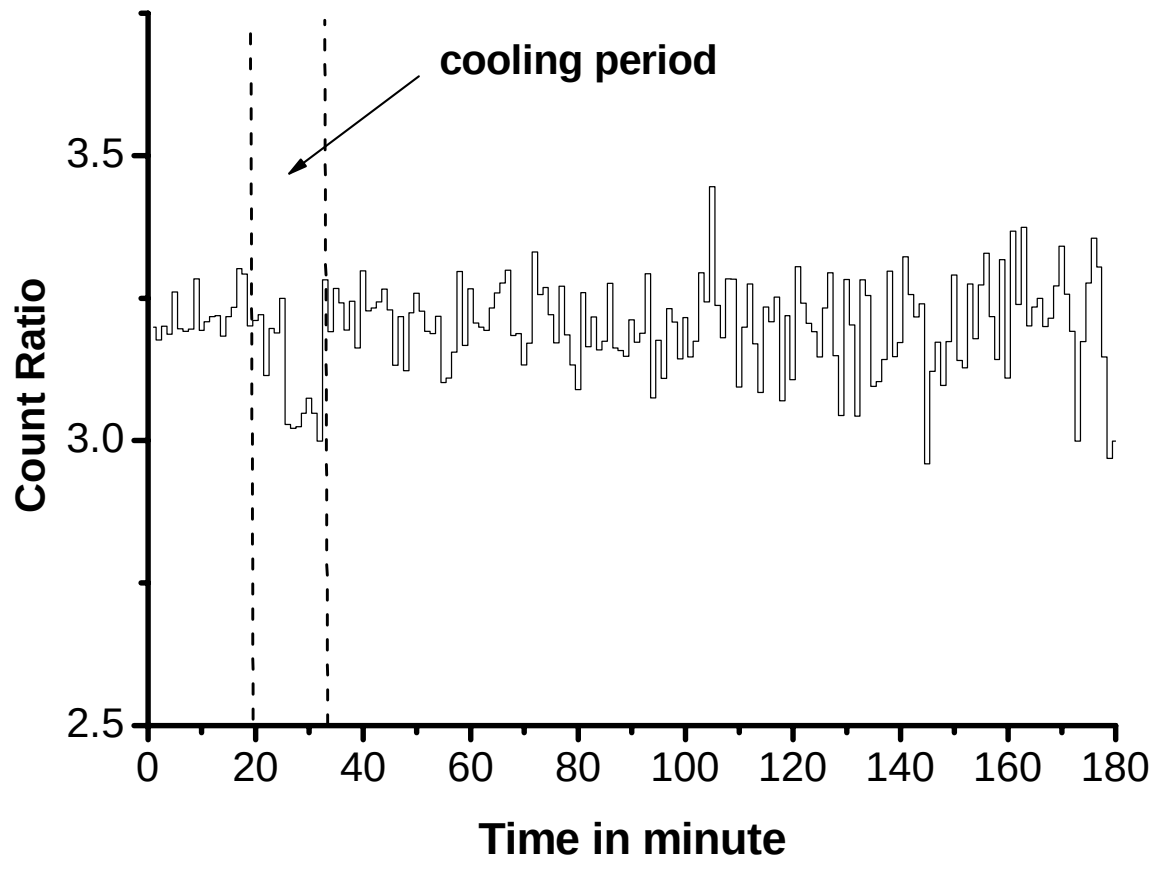

FIG. 8: Time evolution of the Kα/Kβ ratio, which reveals the nitrogen absorption. Each data is collected from 20 channels.

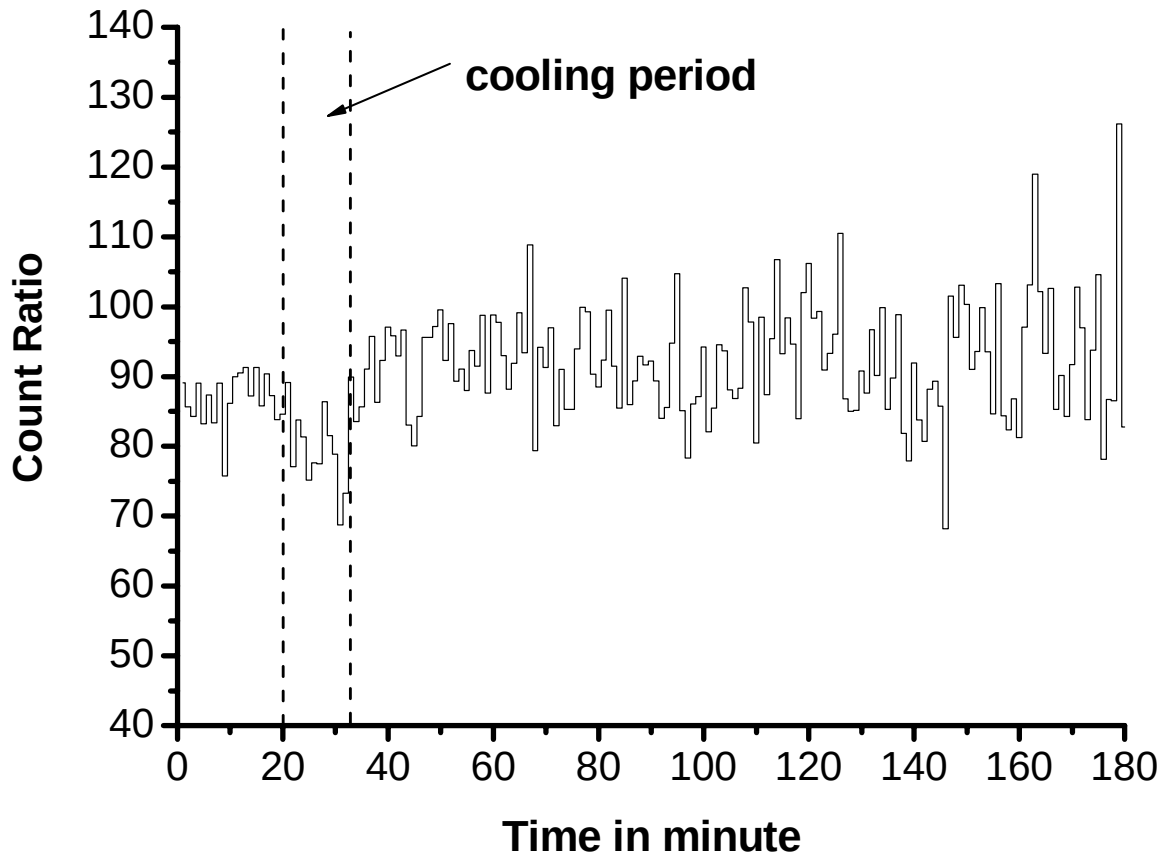

FIG. 9: Time evolution of the internal conversion K/γ, which reveals the K suppression. Each data is collected from 40 channels for K and 16 channels for γ. Thus, the baseline shift is insignificant. The 2% peak pile-up of Kα at the right shoulder of gamma has been removed from the denominator.

## 6. Acknowledgements

We are indebted to Long Wei and Jin Li at institute of High-Energy Physics, Chinese Academy of Sciences for their kindness to lend us the level HPGe detector. We thank Bing Xia for data collection, Mingda Li and Xiangkun Li for data analysis, Hong-Fei Wang for the manuscript preparation, Yanhua Shi, Yexi He and Yinong Liu for fruitful discussion and Yuzheng Lin with his accelerator team, especially Qingxiu Jin and Xiaokui Tao.